\documentstyle[prb,preprint,aps]{revtex}

 \tighten

\def\ra{\rangle}

\def\ver{\arrowvert}

\begin{document}
\noindent
{\large \bf Impossibility of deleting an unknown 
quantum state \hfill}
 
\vskip0.1truein
\noindent
{\bf Arun Kumar Pati and Samuel L.\ Braunstein}
\vskip0.1truein
\noindent
Quantum Optics and Information Group\hfill\break
Informatics, Dean Street, University of Wales\hfill\break
Bangor LL57 1UT, United Kingdom\hfill \break
\noindent
{\vrule width\columnwidth height0.4pt depth0pt\relax}
 
\noindent

{\bf
A photon in an arbitrary polarization state cannot be cloned perfectly 
\cite{wz,dd}. But suppose that at our disposal we have several 
$\pmb{\mbox{\it copies}}$ of an unknown photon. Is it possible to delete
the information content of one or more of these photons by a physical 
process? Specifically, if 
two photons are in the same initial polarization state is there a 
mechanism that produces one photon in the same initial state and the other 
in some standard polarization state. If this can be done, then one would 
create a standard blank state onto which one could copy 
an unknown state approximately, by deterministic cloning \cite{bh,bbhb} 
or exactly, by probabilistic cloning \cite{dg,dg1}. This might be useful 
in quantum computation, where one could store some new information in an 
already computed state by deleting the old information. Here we show that 
the linearity of quantum theory does not allow us to delete a copy of an 
arbitrary quantum state perfectly. Though in a classical computer information 
can be deleted against a copy\cite{rl}, the same task cannot be 
accomplished with quantum information.}

\vskip 1cm

Quantum information has the unique property that it cannot be amplified
accurately. If one could clone an arbitrary state then using non-local 
resources one could send signals faster than light \cite{wz,dd}.
Nevertheless, orthogonal quantum states can be perfectly copied.
Though two non-orthogonal photon-states cannot be copied perfectly by a 
unitary process \cite{hy} they can be copied by a unitary-reduction 
process \cite{dg}. More interestingly, non-orthogonal states from a 
linearly independent set can evolve into a linear superposition
of multiple copy states by a novel cloning machine \cite{akp}.
With the recent advances in quantum information theory such as quantum 
cryptography \cite{bb}, quantum teleportation \cite{bet,tel1,dbet,tel2} 
and quantum computing \cite{dde} we would like to know what we can do with 
the vast amount of information contained in an unknown state and what we
cannot. 

As is now well understood, erasing a single copy of some classical 
information is an irreversible operation. It may only be done with 
some energy cost; a result known as the {\it Landauer erasure 
principle} \cite{rl}. In quantum theory, erasure of a single unknown 
state may be thought of as swapping it with some standard state and 
then dumping it into the environment. The deletion we wish study is 
not the same as irreversible erasure, but is more like reversible 
`uncopying' of an unknown quantum state. We will prove that there is 
no quantum deleting machine which can delete one unknown state 
{\it against\/} a copy in either a reversible or an irreversible manner.

Let us suppose that we have several copies of some unknown information. 
Classically we may delete one copy against the others, uncopying it in 
a perfectly reversible manner \cite{rl}. We shall see that the situation 
is very different in quantum theory. Such a quantum deleting machine 
would involve two initially identical qubits ({\it e.g.}, photons of 
arbitrary polarization) in some state $\ver \psi \ra$ and an ancilla 
in some initial state $\ver A \ra$, which corresponds to the `ready' 
state of the deleting apparatus. The aim of this machine is to delete 
one of two copies of $\ver \psi \ra$ and replace it with some standard 
state of a qubit $\ver\Sigma\ra$. The quantum deleting operation is 
defined for an input $\ver\psi\ra\ver\psi\ra$ such that the linear 
operator acts on the combined Hilbert space of input and ancilla.
That is, it is defined by
\begin{eqnarray}
\ver \psi \ra \ver \psi \ra \ver A \ra  \rightarrow \ver \psi \ra \ver 
\Sigma \ra  \ver A_\psi \ra \;, \label{del2}
\end{eqnarray}
where $\ver A_\psi \ra$ is the final state of the ancilla, which may
in general depend on the polarization of the original photon. (If we 
knew this process was unitary it might work like the time-reverse of 
cloning.) One obvious solution to this equation is to swap the second 
and third states. However, this reduces to the standard erasure 
result where the extra copies have played no role. We will therefore 
explicitly exclude swapping as describing quantum deleting.  

Consider the action of this deleting machine~(\ref{del2}) on a pair of 
horizontally or vertically polarized photons:
\begin{eqnarray}
\ver H \ra \ver H \ra \ver A \ra  &\rightarrow & 
\ver H \ra \ver \Sigma \ra  \ver A_H \ra \nonumber \\
\ver V \ra \ver V \ra \ver A \ra  &\rightarrow&
\ver V \ra \ver \Sigma \ra  \ver A_V \ra \;. \label{tran1}
\end{eqnarray}
We note that the transformation defining our deleting machine~(\ref{del2}) 
does not completely specify its action when the input states are 
non-identical. This is in contrast to the Wootters-Zurek cloning 
transformation \cite{wz} whose definition specifies its action for all
possible inputs. Because of this the transformation corresponding to 
our machine is {\it not\/} the time-reverse of cloning. In fact,
transformation~(\ref{del2}) defines a whole class of possible deleting 
machines which could behave differently if the two inputs are unequal 
or even entangled, {\it e.g.},
\begin{equation}
{1\over\sqrt{2}}(\ver H\ra\ver V\ra+\ver V\ra\ver H\ra) \ver A\ra
\rightarrow \ver \Phi\ra \;, \label{tran2}
\end{equation}
where $\ver \Phi\ra$ might be {\it any\/} state of the combined 
input-ancilla system.

Now for an arbitrary input qubit 
$\ver \psi \ra =  \alpha  \ver H \ra + \beta  \ver V \ra$
(where $\alpha$ and $\beta$ are {\it unknown\/} complex numbers with 
$\ver \alpha \ver^2 + \ver \beta \ver^2 = 1$), linearity and the
transformations~(\ref{tran1}) and~(\ref{tran2}) show that the deleting 
machine yields
\begin{eqnarray}
&&\ver \psi \ra \ver \psi \ra \ver A \ra  \nonumber \\
&=& [\alpha^2 \ver H\ra\ver H \ra + \beta^2 \ver V\ra\ver V \ra +
\alpha \beta ( \ver H\ra\ver V \ra + \ver V\ra\ver H \ra) ]
\ver A \ra \nonumber \\
& \rightarrow& \alpha^2 \ver H \ra \ver \Sigma \ra  \ver A_H \ra
+ \beta^2 \ver V \ra \ver \Sigma \ra  \ver A_V \ra +
\sqrt{2}\alpha \beta\ver\Phi\ra \;, \label{almostfinal}
\end{eqnarray}
which is a quadratic polynomial in $\alpha$ and $\beta$. However, 
if~(\ref{del2}) is to hold, (\ref{almostfinal}) must reduce to
\begin{equation}
(\alpha  \ver H \ra + \beta  \ver V \ra)\ver \Sigma \ra
\ver A_\psi\ra \;, \label{final}
\end{equation}
for all $\alpha$ and $\beta$. Since $\ver\Phi\ra$ is independent of
$\alpha$ and $\beta$ then $\ver A_\psi\ra$ must be linear in them with 
the only solutions being $|\Phi\ra= (\ver H\ra\ver\Sigma\ra\ver A_V\ra+
\ver V\ra\ver\Sigma\ra\ver A_H\ra)/\sqrt{2}$ and
$\ver A_\psi\ra=\alpha  \ver A_H \ra + \beta  \ver A_V \ra$. Further,
since the final state~(\ref{final}) must be normalized for all possible 
$\alpha$ and $\beta$, it follows that the ancilla states $\ver A_H \ra$ and 
$\ver A_V \ra$ are orthogonal. However, as discussed above, (\ref{del2})
is therefore not uncopying at all, but merely swapping onto a 
two-dimensional subspace of the ancilla. It appears that there is no 
option but to move the information around without deleting it. That is, 
the linearity of quantum theory forbids deleting one unknown state against 
a copy. This we call the {\it ``quantum no-deleting'' principle}. This 
principle is complementary in spirit to the no-cloning principle, and 
is expected to play a fundamental role in our future understanding of 
quantum information theory.

We emphasize that copying and deleting of information in a classical 
computer are inevitable operations whereas similar operations cannot 
be realised perfectly in the quantum computers. This has potential 
applications in information processing because it provides {\it intrinsic 
security\/} to quantum files in a quantum computer. No one can obliterate 
a copy of an unknown file from a collection of several copies in a 
quantum computer. In spite of the ``quantum no-deleting'' principle 
one can try to construct a universal and optimal {\it approximate\/}
quantum deleting machine in analogy to optimal quantum cloning 
machines \cite{gm}. When memory in a quantum computer is scarce (at 
least for a finite number of qubits), approximate deleting may play an 
important role in its own way. Though at first glance quantum deleting 
may seem the reverse of quantum cloning, it is not so. 
One can ask is the no-deleting principle 
derivable from the impossibility of superluminal signaling as 
can no-cloning \cite{wz,dd}. Despite the distinction between these
two operations there may be some link between the optimal fidelities of 
approximate deleting and cloning. Nevertheless, we have discovered another 
limitation on quantum information imposed by the linearity of 
quantum theory.

\noindent
{\bf Acknowledgements:} We thank C.\ H.\ Bennett, S.\ Popescu, S.\ Bose, 
L.\ M.\ Duan and N.\ J.\ Cerf for useful comments.\\

\noindent
Correspondence should be addressed to the author. \\
(e-mail: akpati@sees.bangor.ac.uk)\\

Ref: Nature, {\bf 404}, 164 (2000).\\
\end{document}